\documentclass[12pt,a4paper,final]{iopart}

\expandafter\let\csname equation*\endcsname\relax

\expandafter\let\csname endequation*\endcsname\relax
\usepackage{graphicx}
\usepackage{dcolumn}
\usepackage{bm,soul}
\usepackage{color}
\usepackage{ amssymb }
\usepackage{subfigure}   
\usepackage{ amsmath }
\usepackage{ulem}

\begin{document}
 
\title{Large-deviation theory for a Brownian particle on a ring: a WKB approach}

\author{Karel Proesmans}
 \ead{Karel.Proesmans@uhasselt.be}
 \address{Hasselt University, B-3590 Diepenbeek, Belgium.\\
 Coll\`{e}ge de France, PSL, 11 place Marcelin Berthelot, F-75231 Paris Cedex 05, France}
\author{Bernard Derrida}
 \address{
 Coll\`{e}ge de France, PSL, 11 place Marcelin Berthelot, F-75231 Paris Cedex 05, France}

\date{\today}

\begin{abstract}
We study the large deviation function of the displacement of a Brownian particle confined on a ring. In the zero noise limit this large deviation function has a cusp at zero velocity given by the Freidlin-Wentzell theory. We develop a WKB approach to analyse how this cusp is rounded in the weak noise limit.
\end{abstract}


\section{Introduction}
Large deviations have a long history in the mathematical literature \cite{donsker1975asymptotic,ellis1988large,den2008large}{.}
Over the last {decades}, large-deviation theory has become {also} a central part of non-equilibrium statistical mechanics \cite{derrida2007non,touchette_large_2009,bertini2015macroscopic}, in particular in the context of the fluctuation theorem \cite{lebowitz1999gallavotti}.

One of the simplest models to study large-deviation theory is the Brownian particle dragged through a periodic potential \cite{derrida1983velocity,faucheux1995periodic,maes2008steady,chernyak2009non,masharian2018effective}. In the long-time limit, the empirical velocity, $v$, of the Brownian particle satisfies a large-deviation principle:
\begin{eqnarray}
    I(v)=-\lim_{t\rightarrow\infty}\frac{1}{t}\ln P_t(Q_T=vt),
\end{eqnarray}
where $P_t(Q_t)$ is the probability distribution associated with the displacement $Q_t$ after a time $t$.
A fully general expression for this large-deviation function does not exist, and several methods to study it have been constructed, such as variational principles \cite{lacoste2009fluctuation,nemoto2011variational}, tilted dynamics \cite{chetrite2015variational,nyawo2016large}, first-passage time distributions \cite{saito2016waiting}, and underdamped dynamics \cite{fischer2018large}. In the low-noise limit, one can tackle the problem using the Freidlin-Wentzell theory \cite{freuidlin1994random,speck2012large,faggionato2012representation,bouchet2016generalisation,tizon2018effective}. This method is based on the assumption that, in the low-noise limit, the large-deviation function is dominated by the most likely trajectory of the particle, an assumption which, as we wil see, is valid under some conditions. The optimal trajectory can then be found via a Hamiltonian calculation, giving an explicit expression for $I(v)$. 

Near $v=0$, something odd happens; a 'kink' appears in the Freidlin-Wentzell large-deviation function \cite{lebowitz1999gallavotti,nyawo2016large,speck2012large,mehl2008large,baek2015singularities}. Therefore, to get a precise  value of $I(v)$ in this neighbourhood, solving the problem in terms of  an optimal trajectory is no longer sufficient. To overcome this difficulty  we use a tilted generator method \cite{touchette_large_2009}. This method focuses on finding the largest eigenvalue of a Schr\"odinger-like equation, which is generally hard to solve, but {there exist}  methods known from quantum mechanics, such as diffusive Monte-Carlo methods \cite{lecomte2007numerical,PhysRevLett.120.210602} and Rayleigh-Schr\"odinger perturbation theory \cite{baiesi2009computation}, to obtain the lowest eigenvalue. Here, we will solve the equation in the low-but-finite-noise limit using a WKB approach. This approach allows us to understand how the kink of $I(v)$ is rounded in a weak-noise expansion.

We will start  in section \ref{sec1}, by introducing the model and discussing some basic concepts of large-deviation theory. In section \ref{sec2}, we will review the Freidlin-Wentzell approach to derive $I(v)$ and discuss its limitations. In the  main part of this paper  (section \ref{sec3}) we use  a WKB approach  to calculate $I(v)$ or rather its Legendre transform $\mu(\lambda)$ in the case where the force vanishes nowhere  i.e., the case where there is no metastable state. Finally, we end with conclusions and perspectives in section \ref{conc}.

\section{Model \label{sec1}}
\begin{figure}[h]
\begin{center}
\includegraphics[scale=0.35]{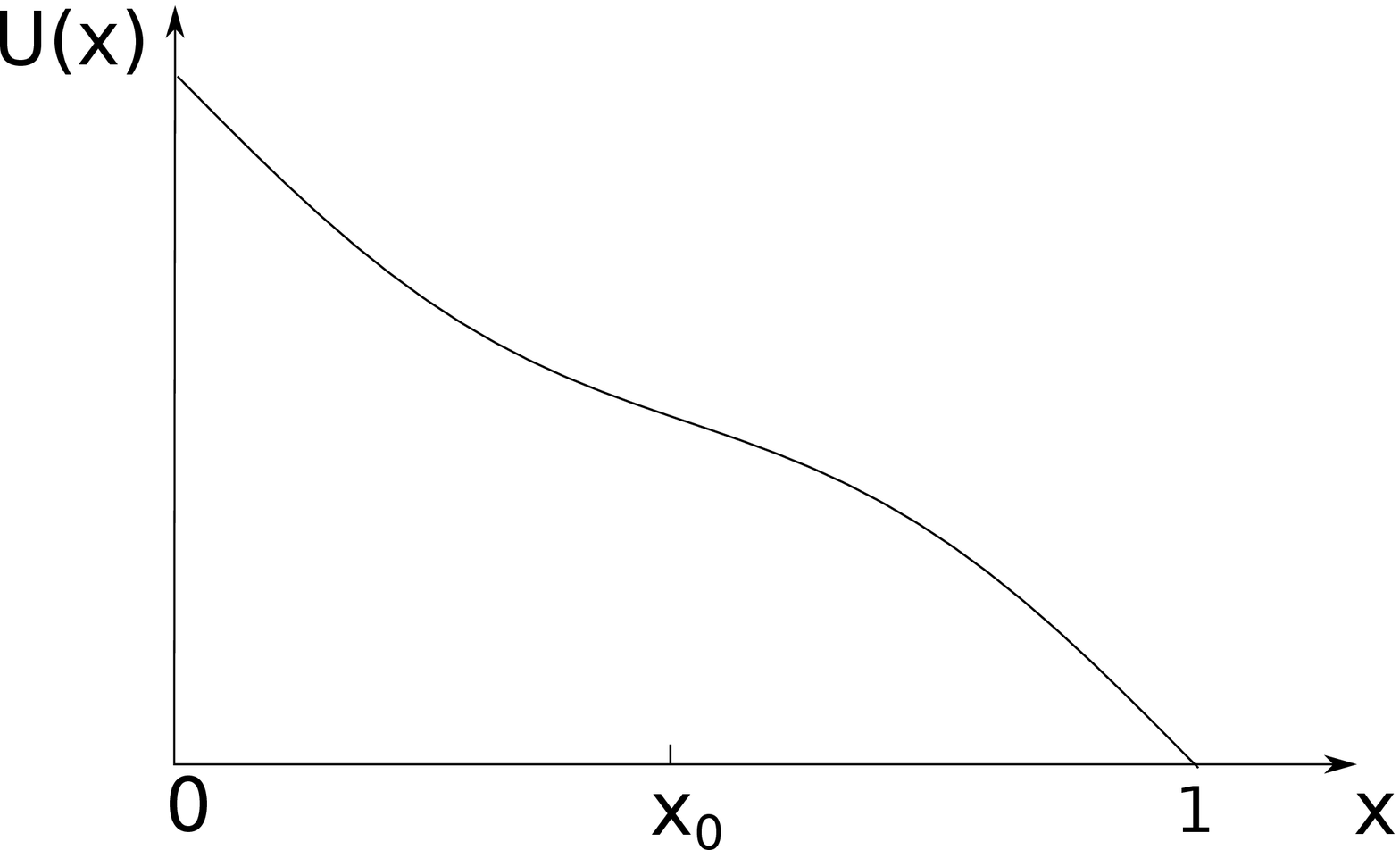}
\includegraphics[scale=0.35]{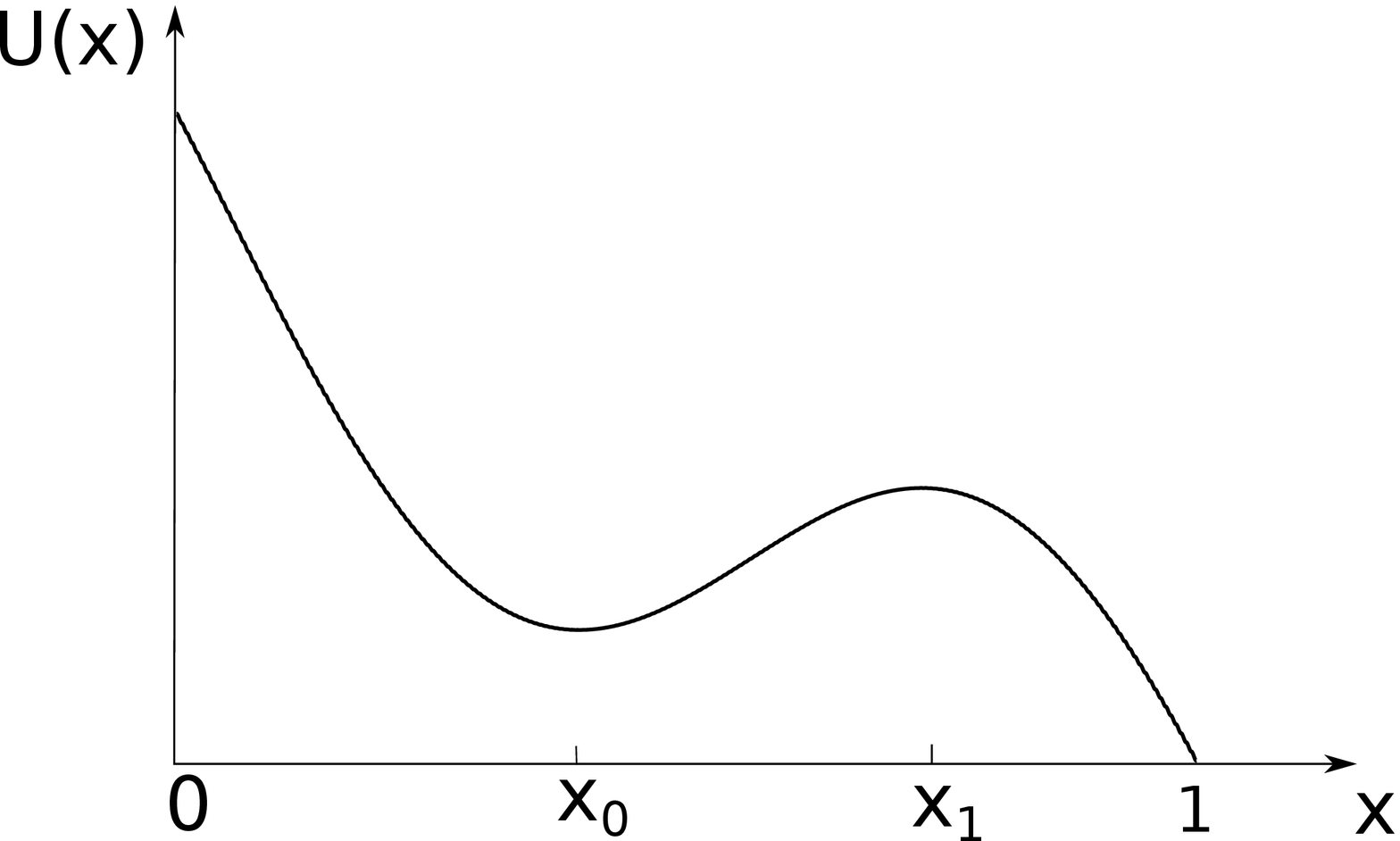}
\end{center}
\caption{One period of $U(x)$. On the left hand side, a case without metastable state. On the right hand side, a case with a single metastable state}
\label{fig:Ux}
\end{figure}
The focus in this paper will be on a Brownian particle dragged through a periodic potential $V(x)$ (period $1$) with a force $f$. For notational simplicity, we shall assume that $f\geq 0$ throughout this text. The particle 'feels' an effective force equal to 
\begin{eqnarray}F(x)=f-V'(x),\qquad V(x+1)=V(x)\end{eqnarray}
 and an associated effective potential 
 \begin{eqnarray}U(x)=V(x)-fx,\end{eqnarray}
 cf.~Fig.~\ref{fig:Ux}. In this section, we shall construct the steady state associated with the position of the particle, and discuss how one can derive the large-deviation function associated with the displacement of the particle. {Throughout this paper,} we will {mainly} focus on periodic potentials with no extrema such as the one drawn{ on the left {panel}} of Fig.~\ref{fig:Ux}.

\subsection{Steady-state distribution}
The position $x(t)$ of the Brownian particle evolves according to an overdamped Langevin equation
\begin{eqnarray}
\dot{x}(t)=-U'(x)+\eta(t),
\end{eqnarray}
where $\eta(t)$ is a Brownian motion,
\begin{eqnarray}
    \left\langle \eta(t)\right\rangle=0,\qquad \left\langle \eta(t)\eta(t')\right\rangle=\epsilon\delta(t-t'),
\end{eqnarray}
and $\epsilon$ is a measure for the strength of the noise.
Associated with this Langevin equation, one can write a Fokker-Planck equation, describing the time-evolution of the probability distribution, $p_t(x)$ associated with $x(t)$:
\begin{eqnarray}
    \frac{d}{dt}p_t(x)=-\frac{d}{dx}\left(F(x)p_t(x)\right)+\frac{\epsilon}{2}\frac{d^2}{dx^2}p_t(x).
\end{eqnarray}
 Therefore, the (time-independent) steady-state distribution, $p_{\textrm{ss}}(x)$ satisfies
\begin{eqnarray}
   -\frac{d}{dx}\left(F(x)p_{\textrm{ss}}(x)\right)+\frac{\epsilon}{2}\frac{d^2}{dx^2}p_{\textrm{ss}}(x)=0.\label{sseq}
\end{eqnarray}
Due to the periodicity, one has 
\begin{eqnarray}p_{\textrm{ss}}(x+1)=p_{\textrm{ss}}(x).\end{eqnarray} This boundary condition fixes the solution of Eq.~(\ref{sseq}):
\begin{eqnarray}
    p_{\textrm{ss}}(x)=&C\exp\left(-\frac{2}{\epsilon}U(x)\right)\nonumber\\&\times\left(\int^x_0dy\exp\left(\frac{2}{\epsilon} U(y)\right)+e^{\frac{2f}{\epsilon}}\int^1_xdy \exp\left(\frac{2}{\epsilon}U(y)\right)\right),\label{pex}
\end{eqnarray}
where $C$ is a normalization constant.
The average velocity of the particle is given by
\begin{eqnarray}
    \left\langle v\right\rangle=\int^1_0dx\, p_{\textrm{ss}}(x)F(x)=\frac{C\epsilon}{2}\left(e^{\frac{2f}{\epsilon}}-1\right).
\end{eqnarray}

In the weak noise limit {($\epsilon$ small)}, the behaviour of the {velocity} depends on the strength of the external force and can be separated in two classes:
\begin{itemize}
    \item If $ f<\max V'(x)$, the effective potential $U(x)$ exhibits a local minimum and maximum, at $x=x_0$ and $x=x_1$ respectively (see {the right panel of} Fig.~\ref{fig:Ux}), leading to a meta-stable state for the particle {at $x=x_0$. The integral of $p_{\textrm{ss}}(x)$ over $x$ is dominated by the neighbourhood of $x_0$ and the second integral in the r.h.s.~of Eq.~(\ref{pex}) is much larger than the first one, so only $x$ in the neighbourhood of $x_0$ and $y$ in the neighbourhood of $x_1$ dominate. This gives
$${C  \, \epsilon \,\pi\exp\left(\frac{2(f+U(x_1)-U(x_0))}{\epsilon}\right) \over \sqrt{-U''(x_0) U''(x_1)}} \simeq 1 \ . $$} 
The particle generally spends most of its time in this  metastable state and the average velocity is exponentially small,
    \begin{eqnarray}
        \left\langle v\right\rangle\simeq \frac{\sqrt{-U''(x_0)U''(x_1)}}{2\pi}e^{\frac{2\left(U(x_0)-U(x_1)\right)}{\epsilon}} \ . \label{vav1}
    \end{eqnarray}
    {Analysing the whole $x$ range} in  Eq.~(\ref{pex}), one can also  see that $p_{\textrm{ss}}(x)$ is exponentially peaked at $x=x_0$, and exhibits a large-deviation principle in terms of $\epsilon$, with non-analytic points {at the values of $x$ where the two terms in Eq.~(\ref{pex}) have the same magnitude \cite{faggionato2012representation,baek2015singularities}.}
    \item If $ f>\max V'(x)$, there are no local minima in the effective potential. Therefore, the probability distribution associated with the position of the particle is much more spread out over the ring and the average velocity of the particle stays finite for arbitrary small noise:
    \begin{eqnarray}
        \left\langle v\right\rangle\simeq \frac{1}{\int_0^1dy F(y)^{-1}}.\label{vav2}
        \end{eqnarray}
    As $f>0$, the second term in Eq.~(\ref{pex}) is dominant, leading to
    \begin{eqnarray}
        p_{\textrm{ss}}(x)\simeq \frac{C'}{F(x)},
\label{Eq13}
    \end{eqnarray}
    where $C'$ again is a  normalisation constant.
\end{itemize}

\subsection{Large-deviation theory}
In the long-time limit, the measured velocity of the Brownian particle will always converge to the average velocity, Eq.~(\ref{vav1})-(\ref{vav2}). All other velocities become exponentially unlikely. This behaviour is described by the associated large-deviation function:
\begin{eqnarray}
    I(v)=-\lim_{t\rightarrow \infty}\frac{1}{t}\ln P_t( Q_t=vt),
\end{eqnarray}
where $Q_t$ is the total displacement of the Brownian particle after time $t$,
\begin{eqnarray}
    Q_t=\int_0^td\tau\, \dot{x}(\tau).
\end{eqnarray}
Furthermore, one can define the cumulant-generating function of the displacement
\begin{eqnarray}
    \mu(\lambda)=\lim_{t\rightarrow\infty}\frac{1}{t}\ln\left\langle e^{t\lambda v}\right\rangle.
\end{eqnarray}
From $\mu(\lambda)$, one can uncover all cumulants associated with the displacement, as the $n$-th derivative of $\mu(\lambda)$ evaluated at $\lambda=0$ is equal to the $n$-th cumulant.
If the cumulant-generating function is strictly convex, one can extract the large-deviation function via a Legendre transform \cite{touchette_large_2009},
\begin{eqnarray}
    I(v)=\max_{\lambda}\left(\lambda v-\mu(\lambda)\right)
\ \ \ \ \ ; \ \ \ \ \ 
    {\mu(\lambda) =\max_{v}\left(\lambda v- I(v))\right)}.\label{LT}
\end{eqnarray}
Therefore, one can determine the large-deviation function by first calculating the cumulant-generating function, $\mu(\lambda)$, and then doing a Legendre transform.

The cumulant-generating function can be found as the largest eigenvalue of a 'tilted' Fokker-Planck operator \cite{touchette2017introduction,derrida2018large}, see also \ref{appldf},
\begin{eqnarray}
    \mu(\lambda)r(x)&=&\lambda F(x)r(x)-\frac{d}{dx}\left(F(x)r(x)\right)\nonumber\\&&+\frac{\epsilon}{2}\left(\lambda^2r(x)-2\lambda\frac{d}{dx}r(x)+\frac{d^2}{dx^2}r(x)\right),
\end{eqnarray}
where $r(x)$ is the associated eigenvector, which satisfies the periodic boundary condition $r(x+1)=r(x)$. {This equation can be simplified by introducing}
\begin{eqnarray}
    s(x)=\exp\left(-\lambda x\right)r(x),\end{eqnarray} leading to
\begin{eqnarray}
    \mu(\lambda)s(x)=-\frac{d}{dx}\left(F(x)s(x)\right)+\frac{\epsilon}{2}\frac{d^2}{dx^2}s(x),\label{REq}
\end{eqnarray}
with boundary condition
\begin{eqnarray}
    s(x+1)=e^{-\lambda}s(x).\label{pbcs}
\end{eqnarray}
In this way, the eigenvalue equation, Eq.~(\ref{REq}), does no longer explicitly depend on $\lambda$, which only appears via the boundary condition, Eq.~(\ref{pbcs}).
As $\mu(\lambda)$ is the largest eigenvalue of a tilted Fokker-Planck operator, it is also the largest eigenvalue of the adjoint operator,
\begin{eqnarray}
    \mu(\lambda)\ell(x)&=&\lambda F(x)\ell(x)+F(x)\frac{d\ell(x)}{dx}\nonumber\\&&+\frac{\epsilon}{2}\left(\lambda^2\ell(x)+2\lambda\frac{d\ell(x)}{dx}+\frac{d^2\ell(x)}{dx^2}\right),
\end{eqnarray}
which can also be simplified by defining $m(x)=\exp\left(\lambda x\right)\ell(x)$:
\begin{eqnarray}
\label{REq1}
    \mu(\lambda) m(x)=F(x)\frac{d}{dx}m(x)+\frac{\epsilon}{2}\frac{d^2}{dx^2}m(x).
\end{eqnarray}
Interestingly, the left and right eigenvector have a physical interpretation
\cite{touchette2017introduction,derrida2018large,chetrite2015nonequilibrium}:
\begin{eqnarray}
    \ell(x)r(x)=m(x)s(x)\sim P(x|v=\mu'(\lambda)).\label{Pcon}
\end{eqnarray}
In words, this means that, up to a normalisation constant, the product of the left and right eigenvector is equal to the probability distribution associated with the position of the particle, conditioned to the average velocity $v=\mu'(\lambda)$

\section{Freidlin-Wentzell theory \label{sec2}}
One way to {try to obtain} the large-deviation function, $I(v)$, is to determine the most likely trajectory leading to the average velocity $v$ \cite{freuidlin1994random,speck2012large,tizon2018effective, derrida2018large}. This approach is based on the assumption that, {in the low noise limit,} the probability associated with a certain value of $v$ is dominated by a single optimal trajectory. The large-deviation function is then given by
\begin{eqnarray}
    I(v)=\lim_{\mathcal{T}\rightarrow\infty}\min_{\left\{ x(t)\right\}}\frac{1}{2\epsilon\mathcal{T}}\int_0^{\mathcal{T}}dt\Big(\dot{x}(t)-F(x(t))\Big)^2,
\end{eqnarray}
with the boundary conditions
\begin{eqnarray}
    x(0)=0,\qquad \frac{Q_\mathcal{T}}{\mathcal{T}}=v,\label{FWBC}
\end{eqnarray}
where one takes the limit $\mathcal{T}\rightarrow \infty$.
The above equation can be solved using Lagrangian techniques:
\begin{eqnarray}
    \dot{x}(t)^2=F(x(t))^2+K,\label{xsol}
\label{EE7}
\end{eqnarray}
where $K$ is an integration constant, which can be determined by the boundary condition, Eqs.~(\ref{FWBC}). Note that, because of Eq.~(\ref{xsol}), the optimal trajectory is periodic, $x(t+v^{-1})=x(t)$. Therefore, one can set $\mathcal{T}=v^{-1}$. This leads to the following expression of $I(v)$ in a parametric form:
\begin{eqnarray}
\label{Iofv}
    I(v)=\frac{v}{\epsilon}\int^1_0dx\left(\frac{2F(x)^2+K}{2\sqrt{F(x)^2+K}}-F(x)\right),
\end{eqnarray}
with
\begin{eqnarray}
    v^{-1}=\int^1_0\frac{dx}{\sqrt{F(x)^2+K}},
\end{eqnarray}
for $v>0$ and
\begin{eqnarray}
I(v)&=&-\frac{v}{\epsilon}\int^1_0dx\left(\frac{2F(x)^2+K}{2\sqrt{F(x)^2+K}}+F(x)\right)
\label{EE0m} \\
v^{-1}&=&-\int^1_0\frac{dx}{\sqrt{F(x)^2+K}}
\label{EE0}
\end{eqnarray}
for $v<0$.
Using Eq.~(\ref{LT}), one can also determine the cumulant generating function:
\begin{eqnarray}
    \lambda=I'(v)=\frac{1}{\epsilon} \int^1_0dx\, \left( {\pm} \sqrt{F(x)^2+K}-F(x)\right),
\label{EE1}
\end{eqnarray}
\begin{eqnarray}
    \mu(\lambda)=\lambda v-I(v)=\frac{K}{2\epsilon}
\label{EE2}
\end{eqnarray}
which gives an implicit equation for $\mu(\lambda)$:
\begin{eqnarray}
    \epsilon\lambda=-f\pm\int^1_0dx\sqrt{2\epsilon\mu(\lambda)+F(x)^2},\label{impcgf}
\label{EE3}
\end{eqnarray}
where the sign associated with the integral{ is everywhere equal to the sign of $v$}.
\begin{figure}[h]
\begin{center}
\includegraphics[scale=0.5]{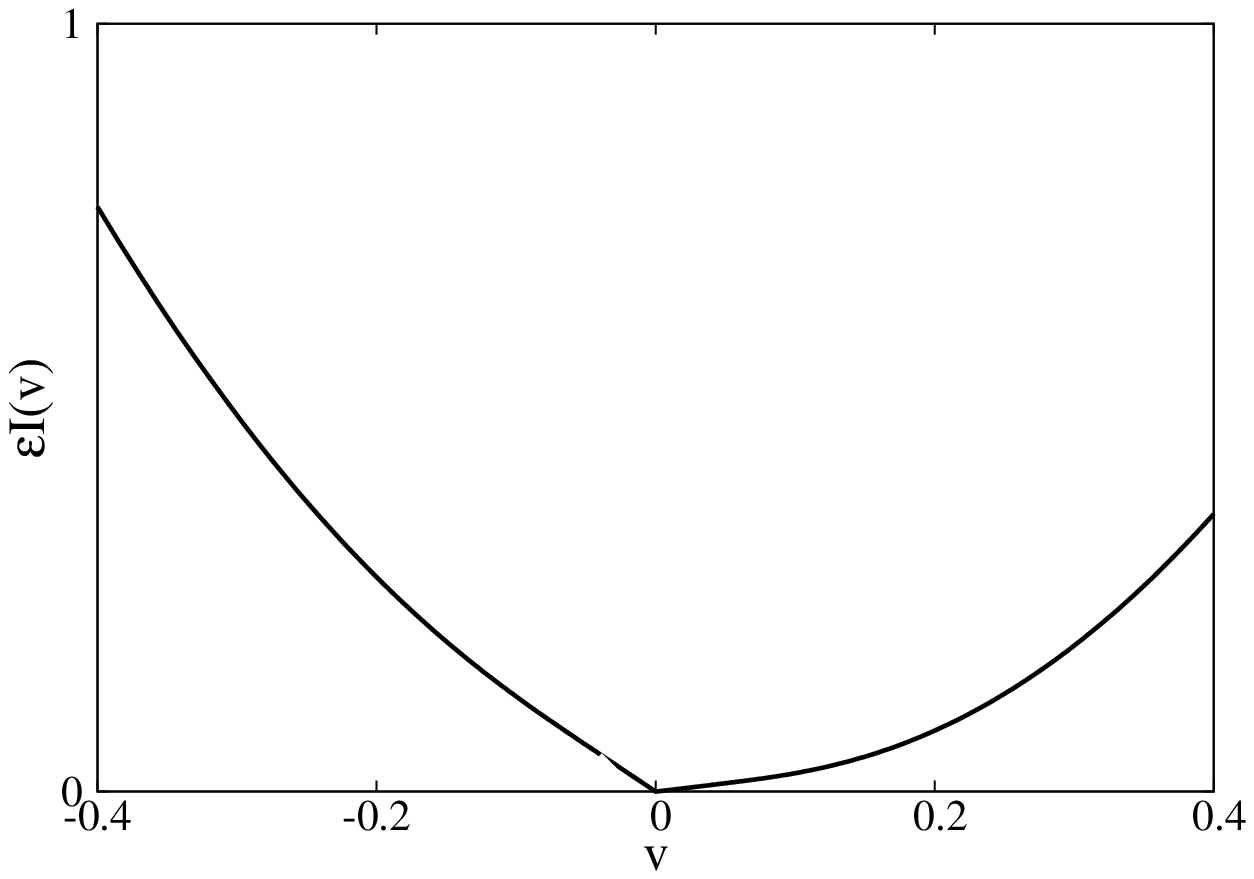}\label{fig:fw1}
\includegraphics[scale=0.5]{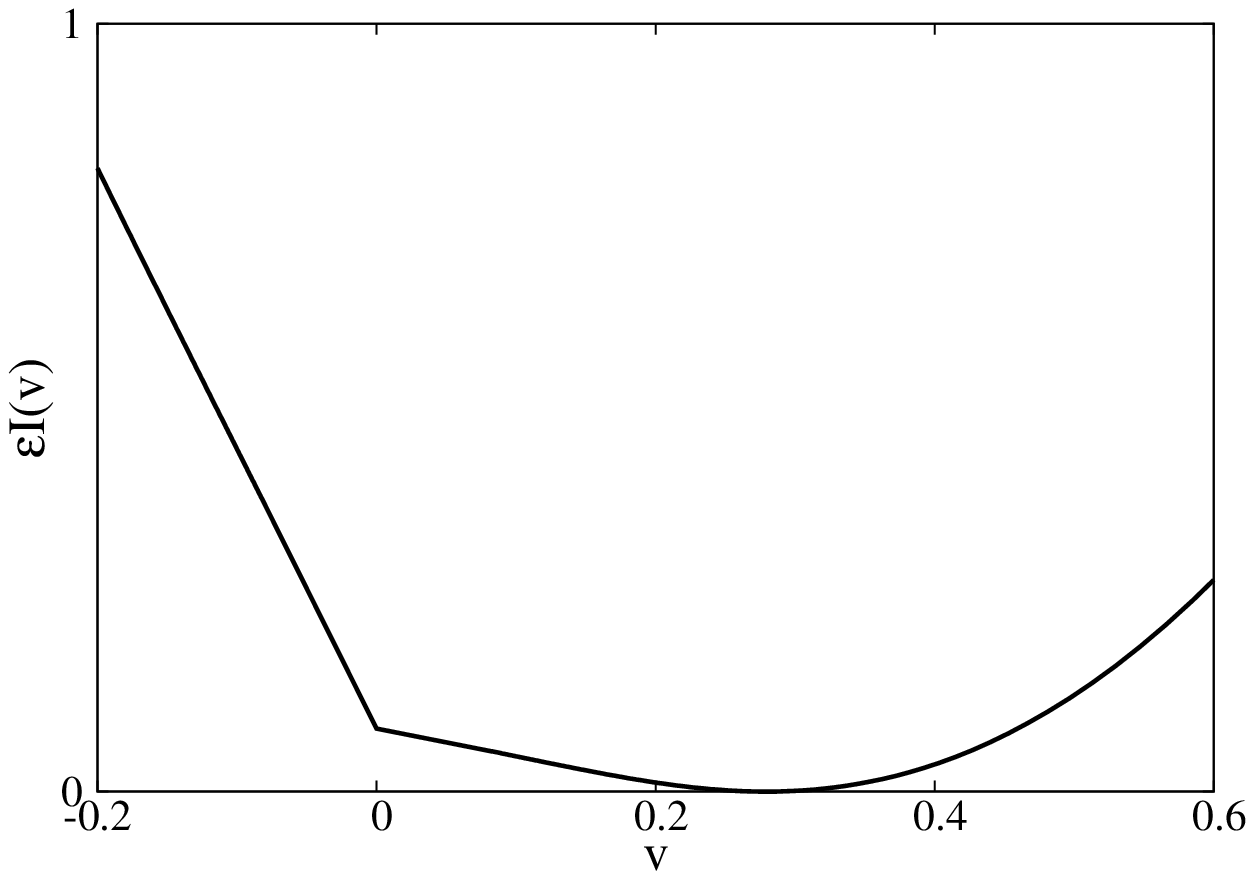}\label{fig:fw2}
\end{center}
\caption{Freidlin-Wentzell large-deviation function with $F(x)=\cos(2\pi x)+f$, with a) $f=1/2$ and b)$f=2$. One sees {in Figure \ref{fig:fw2}} that in the presence of metastable states $I(0)=0$.{ In} both cases a cusp appears at $v=0$.}
\label{fig:FW}
\end{figure}

There is a peculiarity about this solution. {Clearly for the square roots in the above equations  (\ref{Iofv}-\ref{impcgf}) to be defined, one needs that $K \ge -\min_x F(x)^2$ so that
\begin{equation}
\mu(\lambda) \ge -{F(x)^2 \over 2 \epsilon} 
\ \ \ \ \ \text{for all} \ \ x  \ .
\label{mucondition}
\end{equation}
Therefore the above expression (\ref{impcgf}) is only valid outside the following range for $\lambda$ 
}
 \begin{eqnarray}
\label{EE4}
-f-\int^1_0dx\sqrt{F(x)^2-F(x^*)^2}<\epsilon\lambda<-f+\int^1_0dx\sqrt{F(x)^2-F(x^*)^2},\nonumber\\\end{eqnarray}
where $F(x^*)^2$ is the minimal value of $F(x)^2$. If $F(x^*)=0$, i.e., in the presence of  metastable states, this {unreachable range} simplifies to
\begin{eqnarray}
\label{EE5}
    -f-\int^1_0dx\left|F(x)\right|<\epsilon\lambda<-f+\int^1_0dx\left|F(x)\right| \ . 
\end{eqnarray}
This also manifests itself in the large-deviation function, which has a 'cusp' around $v=0$, cf.~Fig.~\ref{fig:FW}. Indeed, one sees {from Eqs.~(\ref{Iofv}-\ref{EE0})  that $K\to - F(x^*)^2$ as $v\to 0$ so that}
\begin{eqnarray}
    I(0^+)=I(0^-)=\frac{F(x^*)^2}{2\epsilon},
\end{eqnarray}
{and from Eq.~(\ref{EE1})}
\begin{eqnarray}
    \epsilon I'(0^-)&=&-f-\int^1_0dx\sqrt{F(x)^2-F(x^*)^2}\nonumber\\&\neq& -f+\int^1_0dx\sqrt{F(x)^2-F(x^*)^2}=\epsilon I'(0^+).
\end{eqnarray}
{To explore the range (\ref{EE4}) or (\ref{EE5})  one needs to study more carefully the limit $\mu \to- {F(x^*)^2 \over 2 \epsilon}$ and this will be done  in the next section  using a WKB approach.}

{It is clear from Eq.~(\ref{EE7}) that $|\dot{x}| = \sqrt{F^2(x)+K}$. The time spent near position $x$ is proportional to $|\dot{x}|^{-1}$.  Therefore the probability $P(x|v)$ of finding the particle in $x$, conditioned on a certain value of the empirical velocity $v$,
is given by
\begin{equation}
\label{EE8}
P(x|v) \simeq {v \over \sqrt{F^2(x)+K}} 
\end{equation}
This of course reduces to Eq.~(\ref{Eq13}) in the limit $\lambda \to 0$ (i.e., $\mu \to 0$ and $K \to 0$).
}

Finally, we note {(see Eqs.~(\ref{Iofv},\ref{EE0m}))} that the large-deviation function satisfies the fluctuation theorem \cite{lebowitz1999gallavotti,gallavotti1995dynamical}:
\begin{eqnarray}
    I(v)=I(-v)-\frac{2v}{\epsilon}\int^1_0dx\, F(x).
\end{eqnarray}

\section{WKB {approach}  when there is no metastable state
\label{sec3}}
In this section, we {obtain $\mu(\lambda)$} by solving the eigenvalue equation, Eq.~(\ref{REq}), in the low-but-finite noise limit. To do this, we look for { an eigenvector, in a WKB form}
\begin{eqnarray}
    s(x)\simeq g(x)\exp\left(\frac{h(x)}{\epsilon}\right),\label{WKBans}
\end{eqnarray}
where {$g(x)$ and $h(x)$ } are unknown functions, independent of $\epsilon$.

{As we expect  from Eq.~(\ref{EE2}) that $\mu(\lambda) = O(\epsilon^{-1})$}
 plugging in Eq.~(\ref{WKBans}) into Eq.~(\ref{REq}) {one gets}:
\begin{eqnarray}
    \frac{2\mu\epsilon+2F(x)h'(x)-h'(x)^2}{2\epsilon}\nonumber\\+\frac{2F'(x)g(x)-g(x)h''(x)+2F(x)g'(x)-2g'(x)h'(x)}{2g(x)}=O(\epsilon).
\end{eqnarray}
Solving this equation gives us a solution for $s(x)$ (up to zero-th order in $\epsilon$ {in the prefactors}):
{
\begin{eqnarray}
    s(x)= \, C_+ \ s_+ (x) \, + \, C_-\ s_-(x)
\label{sWKB}
\end{eqnarray}
where
\begin{eqnarray}
s_\pm(x) =  \sqrt{1 \pm  {F(x) \over \sqrt{2 \epsilon \mu +F(x)^2}}}\exp\left[\frac{1}{\epsilon}\int_0^xdy\,\left(F(y) \pm \sqrt{2\epsilon\mu+F(y)^2}\right)\right] \ \ \ \ \ 
\label{sWKB1}
\end{eqnarray}
Writing that $s(x)$ and its derivatives satisfy 
  the  boundary condition, Eq.~(\ref{pbcs}), implies 
that one of the two constants $C_+$ or $C_-$  vanishes and fixes  the value of $\lambda$
\begin{eqnarray}
    \epsilon\lambda=-\left[f\pm \int^1_0dx\sqrt{2\epsilon\mu(\lambda)+F(x)^2}\right] .\label{cgf0}
\end{eqnarray}
 One recovers that way  the Freidlin-Wentzell result, Eq.~(\ref{impcgf}). 
}

Similarly one can write  the solution of Eq.~(\ref{REq1}) for the left eigenvector in a WKB form 
\begin{eqnarray}
    m(x)= \, C_+' \ m_+ (x) \, + \, C_-'\ m_-(x)
\label{mWKB}
\end{eqnarray}
where
\begin{eqnarray}
m_\pm(x) =  \sqrt{1 \mp  {F(x) \over \sqrt{2 \epsilon \mu +F(x)^2}}}\exp\left[-\frac{1}{\epsilon}\int_0^xdy\,\left(F(y) \pm \sqrt{2\epsilon\mu+F(y)^2}\right)\right] \ \ \ \ \ 
\label{mWKB1}
\end{eqnarray}
{
and again the boundary condition  $m(x+1) = e^{\lambda} m(x)$ forces one of the two constants $C'_+$ or  $C'_-$ to be zero and fixes the value of $\lambda$ as in Eq.~(\ref{cgf0}).
Using  Eq.~(\ref{Pcon})  one  gets for the probability of finding the particle in $x$, conditioned on the velocity $v$ 
\begin{eqnarray}
    P(x|v=\mu'(\lambda))\sim l(x)  r(x)= m(x) s(x)\sim \frac{C}{\sqrt{2\epsilon\mu(\lambda)+F(x)^2}}.
\label{EE9}
\end{eqnarray}
which is exactly what was  obtained in Eq.~(\ref{EE8}) with the Freidlin-Wentzell approach.
Note that in contrast to the left and right eigenvector, Eqs.~(\ref{sWKB})-(\ref{mWKB}), the probability distribution  in Eq.~(\ref{EE9}) does not have any exponential factor, implying that the distribution is not heavily peaked at a certain value, but is relatively spread out over the entire ring.
}

{
All the above calculation 
is valid as long  as  $$\mu  +  {F(x^*)^2 \over  2 \epsilon } = O\left({1} \right)  $$
 $F(x^*)^2=\min_x F(x)^2$. This can be seen as the prefactors in  Eqs.~(\ref{sWKB1}) and (\ref{mWKB1})  diverge in the limit 
$x\to x^*$ and $\epsilon \mu \to   -{F(x^*)^2 \over  2 }$.
}

{
In order to understand this limit, we  consider now the case where $F(x)$ does not vanish on the ring and has a single quadratic minimum $F_0$ at some position $x_0$ 
$$F(x) \simeq F_0 + F_1(x-x_0)^2 + O\Big((x-x_0)^2\Big)$$ and we set 
\begin{equation}
\mu = -{F_0^2 \over  2 \epsilon} + \sqrt{2 F_0 F_1} \left( \nu -{1\over 2} \right)  \label{mu-range} 
\end{equation}
where $\nu -{1\over2} $ is of order $1$  (or smaller) in the limit $\epsilon \to 0$
}

In this range of values of $\mu$, to solve the eigenvalue problem Eq.~(\ref{REq}),  we  decompose the ring into three regions
\begin{itemize}
\item Region I : $0<x < x_0$ and $x_0-x \gg \sqrt{\epsilon} $
\item Region II :  $x_0-x = O( \sqrt{\epsilon})$
\item Region III : $ x_0< x < 1 $ and $x-x_0 \gg \sqrt{\epsilon}$
\end{itemize}
In regions I and III,  one can use solutions analogous to Eqs.~(\ref{sWKB},\ref{sWKB1}) for the eigenvector $s(x)$  solution of Eq.~(\ref{REq}) (see Eq.~(\ref{s1})), whereas in region II  the solution  is of the form 
\begin{equation}
s= \exp \left[ {F_0 (x-x_0) \over \epsilon} + \sqrt{F_0 F_1 \over 2} {(x-x_0)^2 \over \epsilon} \right]
 \ G \left( (2 F_0 F_1)^{1/4} \ {(x-x_0) \over \sqrt{\epsilon} } \right)\label{sol2}
\end{equation} 
where the scaling function $G$ is solution of 
\begin{equation}
\nu \, G = {d\over dz}  (z G) + {1 \over 2} {d^2 G \over dz^2}  \ . 
\label{Herm}
\end{equation}

Our task then is to choose  pairs of  constants $C_+$ and $C_-$  of Eq.~(\ref{sWKB}) in regions I and III and the appropriate solution of the Hermite equation, Eq.~(\ref{Herm}) for the asymptotics of Eq.~(\ref{sol2}) in region II to    match with those of the solutions in regions I and III in the range  $\sqrt{\epsilon} \ll |x-x_0| \ll 1$. 
This is what we do in \ref{WKB-appendix}
where we show that
\begin{equation}
\mu + {F_0^2 \over 2 \epsilon} + {\sqrt{2 F_0 F_1} \over 2} \simeq  {(2 F_0 F_1)^{3 \over 4} \over \sqrt{\epsilon \pi }} \left( e^{\lambda - D_4 + D_2} + e^{-\lambda +D_3-D_1} \right)\label{mufin}
\end{equation}
where the constants $D_1,D_2,D_3,D_4$ are given by Eqs.~(\ref{D1def},\ref{D2def},\ref{D3def},\ref{D4def}).
We see that as $\lambda$ varies in the range  (\ref{EE4}), the $\lambda$-dependence of $ \mu(\lambda)$ is exponentially small.

{At the boundaries $\epsilon\lambda=-\left(f\pm\int^1_0dx\sqrt{F(x)^2-2\epsilon F_0}\right)+O(\epsilon)$, one can connect the two results, Eqs.~(\ref{cgf0}) and (\ref{mufin}), via the formulas
\begin{equation}
-\lambda+D_3-D_1=\ln\left( \frac{2^\nu \sqrt{\pi}}{\Gamma\left(\nu\right)} \left({2 F_0 F_1 \over \epsilon^2} \right)^{{\nu\over 2} -{1 \over 4}}\right),
\end{equation}
near $\epsilon \lambda\approx -f-\int^1_0dx \sqrt{2\epsilon\mu(\lambda)+F(x)^2}$, and
\begin{equation}
\lambda-D_4+D_2=\ln\left( \frac{2^\nu \sqrt{\pi}}{\Gamma\left(\nu\right)} \left({2 F_0 F_1 \over \epsilon^2} \right)^{{\nu\over 2} -{1 \over 4}}\right),
\end{equation}
near $\epsilon \lambda\approx -f+\int^1_0dx \sqrt{2\epsilon\mu(\lambda)+F(x)^2}$. One can check that these equations are in agreement with Eqs.~(\ref{cgf0}) and (\ref{mufin}) in the appropriate limit \cite{tizon2018effective}.
}

\section{Conclusion \label{conc}}
In this paper, we have calculated in the low noise limit the large-deviation and cumulant-generating functions associated with the velocity of a Brownian particle on a ring.  In all cases the large deviation function exhibits a cusp at zero velocity in the limit of zero noise {(see Figures \ref{fig:FW})}. Using  a WKB approach, we could recover the results of the  Freidlin-Wentzell theory at non-zero velocity and analyse the rounding of the cusp in  the weak noise limit. 

{We limited our analysis  to the case of a periodic force with no metastable state.
Our analysis can be extended to the case of one or several metastable states. For example in the case of a single metastable state as in the right panel of figure \ref{fig:Ux}, one would need to consider 5 regions: $x< x_0$, $x$ close to $x_0$, $ x_0 < x < x_1$, $x$ close to $x_1$ and $x_1 < x < 1$ and one would calculate $\mu $ by matching the asymptotics  very much as we did in section \ref{sec3} and \ref{WKB-appendix}.
}

\section*{Acknowledgment}
KP was supported by the Flemish Science Foundation (FWO-Vlaanderen) travel grant V436217N and post-doctoral grant 12J2819N.
{We also thank Bertrand Eynard for very useful discussions on the WKB method.}

\appendix

\section{The large deviation of the current and the deformed Fokker-Planck equation \label{appldf}}

After a short time interval $\Delta t$ one has
\begin{eqnarray}x(t + \Delta t) = x(t)  +F(x(t)) \Delta t + B\end{eqnarray}
and
\begin{eqnarray}Q(t + \Delta t) = Q(t) +F(x(t)) \Delta t + B\end{eqnarray}
where $B$ is a Gausian random variable satisfying
\begin{eqnarray}\langle B \rangle =0 \ \ \ \ \ ; \ \ \ \ \
\langle B^2 \rangle = \epsilon \Delta t \end{eqnarray}

One can write
\begin{eqnarray}P_{t+\Delta t}(x,Q|x_0)& =& \int dQ' \int dx'  \, 
\delta\Big(x-x' - F(x') \Delta t -B\Big) \, \times\nonumber\\&&\qquad\quad\qquad\times
\delta\Big(Q-Q' - F(x') \Delta t -B\Big) \, P_t(x',Q'| x_0)\nonumber\\ \end{eqnarray}
and
\begin{eqnarray}P_{t+\Delta t}(x,Q|x_0) &=& \int dQ' \int dx_0'  \,
\delta\Big(x_0'-x_0 - F(x_0) \Delta t -B\Big) \,\nonumber\\&&\qquad\quad \qquad \times
\delta\Big(Q-Q' - F(x_0) \Delta t -B\Big) \, P_t(x,Q'| x_0') \nonumber\\\end{eqnarray}
and 
therefore 
the  joint probability $P(Q,x|x_0)$  of $Q_t$ and $x_t$ given that  the initial condition $x_0$   evolves according to 
{\small
\begin{eqnarray} {dP(Q,x|x_0) \over dt}  &=& -F(x) {d P(Q,x|x_0) \over dQ}
-{d [F(x) P(Q,x|x_0)]  \over dx} \nonumber\\&&+{\epsilon  \over 2} \left(
{d^2 P(Q,x|x_0) \over d  Q^2}
+2 {d^2 P(Q,x|x_0) \over d Q  \, d x}
+ {d^2 P(Q,x|x_0) \over d x^2}
\right) \ ,
 \end{eqnarray}
and it also satisfies
\begin{eqnarray} {dP(Q,x|x_0) \over dt}  &=& -F(x_0) {d P(Q,x|x_0) \over dQ}
+ F(x_0) {d P(Q,x|x_0)  \over dx_0} \nonumber\\&&+{\epsilon  \over 2} \left(
{d^2 P(Q,x|x_0) \over d  Q^2}
-2 {d^2 P(Q,x|x_0) \over d Q  \, d x_0}
+ {d^2 P(Q,x|x_0) \over d x_0^2}
\right) \ ,
 \end{eqnarray}
}
with the initial condition
\begin{eqnarray}P_0(Q,x|x_0) = \, \delta(Q) \, \delta(x-x_0).\end{eqnarray}

If one introduces the generating function
\begin{eqnarray}\widetilde{P}_t(x|x_0) = \int dQ \, P(Q,x|x_0) \, e^{\lambda Q} \end{eqnarray}
it satisfies
\begin{eqnarray}{d\widetilde{P}(x|x_0) \over dt}  &=& \lambda F(x) \, \widetilde{P}(x|x_0) 
-{d [F(x) \widetilde{P}(x|x_0)]  \over dx} \nonumber\\&&+{\epsilon  \over 2} \left( \lambda^2
\widetilde{P}(x|x_0)
-2 \lambda \,  {d\widetilde{P}(x|x_0) \over   \, d x}
+ {d^2\widetilde{P}(x|x_0) \over d x^2}
\right) \ .
 \end{eqnarray}
and 
\begin{eqnarray}{d\widetilde{P}(x|x_0) \over dt}  = \lambda F(x_0)\,  \widetilde{P}(x|x_0) 
+ F(x_0) {d\widetilde{P}(x|x_0)  \over dx_0} \nonumber\\+{\epsilon  \over 2} \left(
\lambda^2 \, \widetilde{P}(x|x_0) 
+2 \lambda \,  {d\widetilde{P}(x|x_0) \over   \, d x_0}
+ {d^2\widetilde{P}(x|x_0) \over d x_0^2}
\right) \ .
 \end{eqnarray}
with the initial condition
\begin{eqnarray}\widetilde{P}_0(x|x_0) =  \, \delta(x-x_0).\end{eqnarray}

In the long time limit
\begin{eqnarray}\widetilde{P}_0(x|x_0) 
 \sim e^{ \mu(\lambda) \, t } \ r(x) \ \ell(x_0).\end{eqnarray}
 $r(x)$ and $\ell(x)$ are the right and left eigenfunctions  solution of the eigenvalue problem
\begin{eqnarray} \mu(\lambda) \,  r(x)  &=&  \lambda F(x) \,  r(x) 
-{d [F(x) r(x)]  \over dx} \nonumber\\&&+{\epsilon  \over 2} \left(
\lambda^2  r(x)
-2 \lambda {d r(x) \over d x}
+ {d^2 r(x) \over d x^2}
\right) \ ,
 \end{eqnarray}
\begin{eqnarray} \mu(\lambda) \,  \ell(q)  =  \lambda  F(x) \,  \ell(x) 
+ F(x) {d \ell(x)  \over dx} +{\epsilon  \over 2} \left(
\lambda^2  \ell(x)
+2 \lambda {d \ell(x) \over d x}
+ {d^2 \ell(x) \over d x^2}
\right) \ .\nonumber\\
 \end{eqnarray}

\section{The matching of the asymptotics   \label{WKB-appendix}}
In this appendix, we  analyse the situation Eq.~(\ref{mu-range}) and we derive connection formulas between the expressions of
 the solution $s(x)$  in the various regions.
\begin{itemize}
\item In Region I ($0 < x < x_0$) one can write the solution $s(x)$ of Eq.~(\ref{REq}) as
 (see Eqs.~(\ref{sWKB},\ref{sWKB1}))
\begin{align}
s_I(x)= &   c_1\  \sqrt{  {F(x) \over  \sqrt{ F(x)^2-F_0^2}}+1}   \
\exp\left[ \int_0^x dy \left({ F(y) + \sqrt{ F(y)^2-F_0^2} \over \epsilon} + {(\nu-{1\over 2} )  \sqrt{2 F_0 F_1}  \over \sqrt{F(y)^2-F_0^2}} \right) \right]
\nonumber
\\
    +  & c_2 \ \sqrt{  {F(x) \over  \sqrt{ F(x)^2-F_0^2}}-1}   \
\exp\left[ \int_0^x dy \left({ F(y) - \sqrt{ F(y)^2-F_0^2} \over \epsilon} - {(\nu-{1\over 2} )  \sqrt{2 F_0 F_1}  \over \sqrt{F(y)^2-F_0^2}} \right) \right]
\label{s1}
\end{align}
For  $x\to x_0$ in this region I  this leads to  the following asymptotics 
and

\begin{align}
s_I(x) \simeq   
\left( {F_0\over 2 F_1}\right)^{1\over 4} 
 & \left(   
 c_1 \   (x_0-x)^{-\nu} 
\exp\left[ D_1 -{F_0(x_0-x) \over \epsilon} - \sqrt{F_0 F_1 \over 2} {(x_0-x)^2 \over  \epsilon}   \right]
\right.
\nonumber 
 \\
 &   \left. + c_2\  (x_0-x)^{\nu-1}  \exp\left[ D_2 -{F_0(x_0-x) \over \epsilon} + \sqrt{F_0 F_1 \over 2} {(x_0-x)^2 \over  \epsilon}  \right] 
\right) 
\label{As1}
\end{align}
where
\begin{equation}D_1 = \left(\nu -{1\over2} \right) \log x_0 + \int_0^{x_0} dy \left({ F(y) + \sqrt{ F(y)^2-F_0^2} \over \epsilon} 
+ { \left(\nu -{1\over2} \right) \sqrt{2 F_0 F_1} \over \sqrt{F(y)^2-F_0^2}} 
- {\nu -{1\over2}  \over x_0-y} \right) \label{D1def}
\end{equation}
and  
\begin{equation}D_2 = -\left(\nu -{1\over2} \right) \log x_0 + \int_0^{x_0} dy \left({ F(y) - \sqrt{ F(y)^2-F_0^2} \over \epsilon} 
- { \left(\nu -{1\over2} \right) \sqrt{2 F_0 F_1} \over \sqrt{F(y)^2-F_0^2}} 
+ { \nu -{1\over2}  \over x_0-y} \right). \label{D2def}
\end{equation}

\item Similarly in Region III ($x_0 < x < 1$)
\begin{align}
s_{III}(x)= & c_3  \ \sqrt{  {F(x) \over  \sqrt{ F(x)^2-F_0^2}}+1}   \
\exp\left[ -\int_x^1 dy \left({ F(y) + \sqrt{ F(y)^2-F_0^2} \over \epsilon} + {(\nu-{1\over 2} )  \sqrt{2 F_0 F_1}  \over \sqrt{F(y)^2-F_0^2}} \right) \right]
\nonumber
\\
    + & c_4  \ \sqrt{  {F(x) \over  \sqrt{ F(x)^2-F_0^2}}-1}   \
\exp\left[ -\int_x^1 dy \left({ F(y) - \sqrt{ F(y)^2-F_0^2} \over \epsilon} - {(\nu-{1\over 2} )  \sqrt{2 F_0 F_1}  \over \sqrt{F(y)^2-F_0^2}} \right) \right]
\label{s3}
\end{align}
which gives as $x \to x_0$

\begin{align}
s_{III}(x) \simeq  
\left( {F_0\over 2 F_1}\right)^{1\over 4} 
 & \left(   
 c_3 \   (x-x_0)^{\nu-1} 
\exp\left[ D_3 +{F_0(x-x_0) \over \epsilon} + \sqrt{F_0 F_1 \over 2} {(x_0-x)^2 \over  \epsilon}   \right]
\right.
\nonumber 
 \\
 &   \left. + c_4\  (x_0-x)^{-\nu}  \exp\left[ D_4 +{F_0(x-x_0) \over \epsilon} - \sqrt{F_0 F_1 \over 2} {(x_0-x)^2 \over  \epsilon}  \right] 
\right) 
\label{As3}
\end{align}
where 
\begin{equation}D_3 = -\left(\nu -{1\over2} \right) \log (1-x_0) - \int_{x_0}^1 dy \left({ F(y) + \sqrt{ F(y)^2-F_0^2} \over \epsilon} 
+ { \left(\nu -{1\over2} \right) \sqrt{2 F_0 F_1} \over \sqrt{F(y)^2-F_0^2}} 
- {\nu -{1\over2}  \over y-x_0} \right) \label{D3def}
\end{equation}
and 
\begin{equation}D_4 = \left(\nu -{1\over2} \right) \log (1-x_0) - \int_{x_0}^1 dy \left({ F(y) - \sqrt{ F(y)^2-F_0^2} \over \epsilon}
- { \left(\nu -{1\over2} \right) \sqrt{2 F_0 F_1} \over \sqrt{F(y)^2-F_0^2}}
+ { \nu -{1\over2}  \over y- x_0} \right).\label{D4def}             
\end{equation}
\item Finally in Region II ($x-x_0 = O( \sqrt{\epsilon})$)  the solution is of the form Eq.~(\ref{sol2})
with the following asymptotics (see Eqs.~(\ref{C2},\ref{C3})):
\\
for $(x-x_0)/\sqrt{\epsilon} \to - \infty$
\begin{align}
s_{II} \simeq \exp \left[ {F_0 (x-x_0) \over \epsilon}\right] 
&   \left( V 
  \left({2 F_0 F_1 \over \epsilon^2} \right)^{\nu-1 \over 4}
(x_0-x)^{\nu-1}
 \exp \left[ \sqrt{F_0 F_1 \over 2} {(x-x_0)^2 \over \epsilon} \right]
\right.
\nonumber  \\
 &  \left. + W 
\left({2 F_0 F_1 \over \epsilon^2} \right)^{-{\nu \over 4}} 
 (x_0-x)^{-\nu}  
 \exp \left[ -\sqrt{F_0 F_1 \over 2} {(x-x_0)^2 \over \epsilon} \right]
 \right)
\label{As2a}
\end{align}
and for  $(x-x_0)/\sqrt{\epsilon} \to + \infty$
\begin{align}
s_{II} \simeq \exp \left[ {F_0 (x-x_0) \over \epsilon}\right] 
&   \left( V'
  \left({2 F_0 F_1 \over \epsilon^2} \right)^{\nu-1 \over 4}
 (x-x_0)^{\nu-1}
 \exp \left[ \sqrt{F_0 F_1 \over 2} {(x-x_0)^2 \over \epsilon} \right]
\nonumber
\right. \\
 &  \left. + \  W'
  (x-x_0)^{-\nu}  
 \left( {2 F_0 F_1 \over \epsilon^2} \right)^{-{\nu \over 4}} 
 \exp \left[ -\sqrt{F_0 F_1 \over 2} {(x-x_0)^2 \over \epsilon} \right]
 \right)
\label{As2b}.
\end{align}
\end{itemize}

Now 
using   the boundary condition Eq.~(\ref{pbcs})  one has
\begin{equation}
\label{c3-c4}
c_3 = c_1 e^{-\lambda} 
\ \ \ \ \ ; \ \ \ \ 
c_4 = c_2 e^{-\lambda} 
\end{equation}
and matching the asymptotics, on the one hand Eqs.~(\ref{As1}) and (\ref{As2a}) and on the other hand Eqs.~(\ref{As3}) and (\ref{As2b}), one gets using Eq.~(\ref{C8}) that $\lambda$ should satisfy
\begin{equation}
\label{res1a}
e^{2 \lambda} - e^{ \lambda} \left(  {X(\nu) \over \Gamma(\nu)}  e^{D_4-D_2}  +  {\Gamma(\nu) (1-Z^2) \over X(\nu)} e^{D_3-D_1} \right)  + e^{D_4+D_3-D_1-D_2} =0 \end{equation} 
where
\begin{equation}
 X(\nu)= 2^\nu \sqrt{\pi} \left({2 F_0 F_1 \over \epsilon^2} \right)^{{\nu\over 2} -{1 \over 4}}.
\end{equation}

{For $\epsilon$ small, one has $D_4-D_2 \gg D_3-D_1 $. This, combined with $Z=1+O(\nu)$ and $\nu\ll \epsilon^{-1}$ one can see that the term containing $Z$ becomes negligible over the entire range the range (\ref{EE4}), i.e., 
$$D_3-D_1 < \lambda < D_4-D_2.$$
This simplifies the above equation to
\begin{equation}
    \Gamma(\nu)=\frac{X(\nu)}{e^{\lambda-D_4+D_2}+e^{-\lambda+D_3-D_1}}.\label{nusol1}
\end{equation}
Generally, $\lambda-D_4+D_2$ and $-\lambda+D_3-D_1$ are of the order $\epsilon^{-1}$, and in this regime the above equation can only be satisfied for $\nu\ll 1$, leading to
\begin{equation}
\nu =\frac{\left(2F_0F_1\right)^{\frac{1}{4}}}{\sqrt{\pi\epsilon}}\left( e^{\lambda - D_4 + D_2} + e^{-\lambda +D_3-D_1} \right).
\label{nusol}
\end{equation}
One can see that for $\lambda\approx D_3-D_1$ or $\lambda\approx D_4-D_2$ this simplification does no longer hold. In these regimes, one gets
\begin{equation}
-\lambda+D_3-D_1=\ln\left(\frac{X(\nu)}{\Gamma(\nu)}\right).
\end{equation}
and
\begin{equation}
\lambda-D_4+D_2=\ln\left(\frac{X(\nu)}{\Gamma(\nu)}\right),
\end{equation}
respectively. For $\nu\gg 1$ this simplifies to
\begin{equation}
    \lambda+\frac{\int_0^1dy (F(y)+\sqrt{F(y)^2-F_0^2})}{\epsilon}=\nu\log\left(\nu\epsilon\right)
\end{equation}
and
\begin{equation}
    \lambda-\frac{\int_0^1dy (F(y)-\sqrt{F(y)^2-F_0^2})}{\epsilon}=-\nu\log\left(\nu\epsilon\right).
\end{equation}
This result can be verified by taking the limit to the boundary of Eqs.~(\ref{cgf0}), which leads to exactly the same result \cite{tizon2018effective}.
}

\section{On the asymptotics of the solution of Eq.~(\ref{Herm}) \label{Hermite}}

In this appendix we discuss some aspects of the connection formula of the asymptotics at $z\to + \infty$ and at $z \to - \infty$ of a solution $G$ of
\begin{equation}
\nu \, G = {d\over dz}  (z G) + {1 \over 2} {d^2 G \over dz^2} 
\label{C1}
\end{equation}

For large $z$ one expects either
$G \sim z^{\nu-1} $ or $G \sim e^{-z^2} z^{-\nu}$
 and our goal is  to relate   between  the pair $V, W$ to  the pair $V', W'$ which characterize the asymptotics at $\pm \infty$ 

\begin{equation}
G \simeq V  (-z)^{\nu-1} \Big( 1 + {(\nu-1)(\nu-2) \over 4 \, z^2} + \cdots \Big) 
\  +  \ W \  {e^{-z^2} \over (-z)^\nu}  \Big( 1 - {\nu(\nu+1) \over 4 \, z^2} + \cdots \Big) 
\ \ \ \ \ \text{as} \ \ \ \ \ z \to - \infty
\label{C2}
\end{equation}
and
\begin{equation}
G \simeq V'  z^{\nu-1} \Big( 1 + {(\nu-1)(\nu-2) \over 4 \, z^2} + \cdots \Big) 
\  +  \ W'  \ {e^{-z^2} \over z^\nu}  \Big( 1 - {\nu(\nu+1) \over 4 \, z^2} + \cdots \Big) 
\ \ \ \ \ \text{as} \ \ \ \ \ z \to + \infty
\label{C3}
\end{equation}
The goal of this appendix is to show  that  
\begin{equation}
V' =- Z \,  V + {2^\nu \sqrt{\pi}  \over \Gamma(\nu)} W 
\ \ \ \ \ ; \ \ \ \ \ 
W'  ={(1-Z^2) \, \Gamma(\nu) \over 2^\nu \sqrt{\pi}} V + Z \,  W 
\label{C8}
\end{equation}
where
\begin{equation}
\label{C9}
Z=\cos (\pi \nu).
\end{equation}

By expanding around $z=0$, a general solution of Eq.~(\ref{C1})
can be written as
\begin{equation}
G= g  \,G_3 + g' \, G_4
\label{C4}
\end{equation}
 where $G_3$ and  $G_4$ 
are the even and the odd solutions
$$G_3  =  
 \sum_{n \ge 0} (-)^n z^{2n} {\Gamma(2 n - \nu) \, \Gamma(-{\nu \over 2}) \over  \Gamma(-\nu) \, \Gamma(n-{\nu \over 2}) \, (2n)!}  \ = \ 
1 + (\nu-1) z^2 + {(\nu-1)(\nu-3) \over 6} z^4 + \cdots   $$
and
$$G_4= 
\sum_{n \ge 0} (-)^n z^{2n+1} { 2^{2 n} \, \Gamma( n+1 - {\nu\over 2} )  \over   \Gamma(1-{\nu \over 2}) \, (2n+1)!}  \ = \ 
  z +{ \nu-2\over 3} z^3 + {(\nu-2)(\nu-4) \over 30} z^5 + \cdots   $$
\\ \ \\
If one defines (assuming that $\nu$ is not an integer or half an integer)
$$G_1 = \int_{-\infty + i 0}^\infty e^{-z^2+ t z - {t^2 \over 4}} \, t^{\nu-1} dt $$
$$G_2 = \int_{-\infty - i 0}^\infty e^{-z^2+ t z - {t^2 \over 4}}\,  t^{\nu-1} dt $$
one has $$g_1 = \Gamma\left({\nu \over 2} \right)
 \left(1- e^{i \pi \nu} \right) 2^{\nu-1}
\ \ \ \ \ ; \ \ \ \ \ 
g_1' = \Gamma\left({\nu +1 \over 2} \right) \left(1+ e^{i \pi \nu} \right) 2^{\nu}
$$
and
 $$g_2 = \Gamma\left({\nu  \over 2} \right) \left(1- e^{-i \pi \nu} \right) 2^{\nu-1}
\ \ \ \ \ ; \ \ \ \ \ 
g_2' = \Gamma\left({\nu +1 \over 2} \right) \left(1+ e^{-i \pi \nu} \right) 2^{\nu}
$$
Therefore
\begin{equation}
G_3 ={1 \over 2^\nu \, \Gamma\left({\nu \over 2} \right)\, (1- e^{i \pi \nu})} \Big(G_1- e^{i \pi \nu} G_2\Big) 
\label{C5}
\end{equation}
\begin{equation}
G_4 ={1 \over 2^{\nu+1}  \, \Gamma\left({\nu+1 \over 2} \right)\, (1+ e^{i \pi \nu})} \Big(G_1+ e^{i \pi \nu}G_2 \Big). 
\label{C6}
\end{equation}

Because
\begin{equation}
G_2-G_1 = \int_{-\infty - i 0}^{-\infty+i 0}  e^{-z^2+ t z - {t^2 \over 4}} t^{\nu-1} dt 
\label{FF1}
\end{equation}
and because this integral is dominated for large $z$ by the neighborhood of $t=0$ one has the following asymptotics for $z \to + \infty$

\begin{equation}
G_2-G_1 \sim \left(e^{i \pi \nu} - e^{-i \pi \nu} \right)\ e^{-z^2}\   \left( 
{\Gamma(\nu)  \over z^\nu}   -
{\Gamma(\nu+2)  \over 4 z^{\nu+2}}  + \cdots \right)
\label{C6a}
\end{equation}
On the other hand for large positive $z$ a saddle point calculation leads to
$$G_1 \sim G_2 \sim 2^\nu \sqrt{\pi} z^{\nu-1} \left( 1 + {(\nu-1) (\nu-2) \over 4 z^2} + \cdots \right). $$
Then from Eqs.~(\ref{C5},\ref{C6}) one gets  for large positive $z$ 
\begin{equation}
G_3 \simeq     { \sqrt{\pi} \over  \Gamma({\nu \over 2})} \,  z^{\nu-1}  \ \ \ \ \ ; \ \ \ \ \
G_4 \simeq   { \sqrt{\pi} \over 2  \, \Gamma({\nu +1 \over 2})} \,  z^{\nu-1} 
\label{C7}
\end{equation}
and  from Eq.~(\ref{FF1})
 $$ 
 \Gamma\left({\nu \over 2}\right) 2^{\nu-1} G_3 
-2^\nu \, \Gamma\left({\nu +1 \over 2}\right) G_4 
\simeq \Gamma(\nu) {e^{-z^2} \over z^\nu} $$
So if one postulates that for  $z \to + \infty$
\begin{equation}
G_3= {\sqrt{\pi} \over  \Gamma({\nu \over 2} )} \left[  z^{\nu-1} \left( 1 + {(\nu-1) (\nu-2) \over 4 z^2} + \cdots \right) + {\beta  \over 2^{\nu-1} }{e^{-z^2} \over z^\nu }\left( 1+  \cdots  \right)  \right] 
\label{FF3}
\end{equation}
\begin{equation}
G_4= {\sqrt{\pi} \over 2 \Gamma({\nu +1 \over 2} )} \left[  z^{\nu-1} \left( 1 + {(\nu-1) (\nu-2) \over 4 z^2} + \cdots \right) + {\gamma  \over 2^{\nu-1} }{e^{-z^2} \over z^\nu }\left( 1+  \cdots  \right)  \right] 
\label{FF4}
\end{equation}
 one should have
\begin{equation}
\beta - \gamma = {1\over \sqrt{\pi}} \Gamma(\nu) 
\label{betagamma}
\end{equation}
In the above expressions  $\beta$ and $\gamma $ are factors of subdominant terms   and they are a priori ill defined unless one  specifies how the dominant divergent series is resummed.

A  general  solution of Eq.~(\ref{C1}) can always be written as 
$$G= x \, G_3 + y \, G_4$$
Then  one has  (see Eqs.~(\ref{FF3},\ref{FF4}))
$$V= { \sqrt{\pi} \over  \Gamma({\nu \over 2} ) } x
- { \sqrt{\pi} \over 2 \Gamma({\nu+1 \over 2} ) } y
\ \ \ \ \ ; \ \ \ \ \
W= {2 \sqrt{\pi} \over 2^\nu \Gamma({\nu \over 2} ) } x  \, \beta
- {2 \sqrt{\pi} \over 2^{\nu+1} \Gamma({\nu+1 \over 2} ) } y \, \gamma
$$

$$V'= { \sqrt{\pi} \over  \Gamma({\nu \over 2} ) } x
+ { \sqrt{\pi} \over 2 \Gamma({\nu+1 \over 2} ) } y
\ \ \ \ \ ; \ \ \ \ \
W'= {2 \sqrt{\pi} \over 2^\nu \Gamma({\nu \over 2} ) } x \, \beta
+ {2 \sqrt{\pi} \over 2^{\nu+1} \Gamma({\nu+1 \over 2} ) } y \, \gamma.
$$
Eliminating $x$ and $y$ one gets Eq.~(\ref{C8})
where 
\begin{equation}
Z= {\sqrt{\pi} \over \Gamma(\mu)} (\beta+ \gamma) .
\label{FF5}
\end{equation}

So far $Z$ is  undetermined, and as mentionned earlier it depends on the way the dominant contribution is resummed in Eqs.~(\ref{FF3},\ref{FF4}). This is related to Stokes phenomenon \cite{temme2015asymptotic}.

As for real positive $z$ the solutions $G_1$ and $G_2$ are complex conjugates one can consider that their real part is by definition the  resummed dominant contribution of the large $z$ asymptotics. This implies (see Eq.~(\ref{C6a}))
$$G_2 \simeq 2 \sqrt{\pi}  z^{\nu-1} \Big(1 + \cdots  \Big) \ \ +  \ \ {\Gamma(\nu) \over 2} \left(e^{i \pi \mu} - e^{- i \pi \nu} \right)  {e^{-z^2} \over z^\nu}$$
$$G_1 \simeq 2 \sqrt{\pi}  z^{\nu-1} \Big( 1 + \cdots \Big) \ \  -\ \  {\Gamma(\nu) \over 2 }\left(e^{i \pi \nu} - e^{- i \pi \nu} \right)  {e^{-z^2} \over z^\nu}.$$
This gives Eqs.~(\ref{C5},\ref{C6},\ref{FF3},\ref{FF4})
$$\beta= {\Gamma(\nu) \Big( 1+ \cos(\pi \nu)\Big) \over 2 \sqrt{\pi}} \ \ \ \ ; \ \ \ \ 
\gamma = {\Gamma(\nu) \Big(-1+ \cos(\pi \nu)\Big) \over 2 \sqrt{\pi}} $$
so that (see Eq.~(\ref{FF5}))
$$Z= \cos(\pi \nu)$$  as in  Eq.~(\ref{C9}).

\section*{Bibliography}
\providecommand{\newblock}{}

\end{document}